\documentclass[12pt]{article}

\usepackage{scicite}

\usepackage{times}
\usepackage{url}
\usepackage{float}
\restylefloat{table}
\usepackage{hyperref}
\usepackage{bigstrut}
\usepackage{listings}
\usepackage{xcolor}
\usepackage{graphicx}

\definecolor{codegreen}{rgb}{0,0.6,0}
\definecolor{codegray}{rgb}{0.5,0.5,0.5}
\definecolor{codepurple}{rgb}{0.58,0,0.82}
\definecolor{backcolour}{rgb}{0.95,0.95,0.92}

\lstdefinestyle{mystyle}{
    backgroundcolor=\color{backcolour},   
    commentstyle=\color{codegreen},
    keywordstyle=\color{magenta},
    numberstyle=\tiny\color{codegray},
    stringstyle=\color{codepurple},
    basicstyle=\ttfamily\tiny
,
    breakatwhitespace=false,         
    breaklines=true,                 
    captionpos=b,                    
    keepspaces=true,                 
    numbers=left,                    
    numbersep=5pt,                  
    showspaces=false,                
    showstringspaces=false,
    showtabs=false,                  
    tabsize=2
}

\newenvironment{sciabstract}{%
\begin{quote} \bf}
{\end{quote}}

\title{Linear Approximate Pattern Matching Algorithm}

\author{Anas Al-okaily,$^{1\ast}$ Abdelghani Tbakhi$^{1\ast}$ \\
\\
\normalsize{$^{1}$Department of Cell Therapy \& Applied Genomics, King Hussein Cancer Center} \\
\normalsize{Amman, 11941, Jordan.}
\\
\\
\normalsize{$^\ast$To whom correspondence should be addressed; E-mail:  AA.12682@khcc.jo}
\\
}

\date{}

\begin{document} 


\baselineskip24pt


\maketitle 

\begin{sciabstract}

Pattern matching is a fundamental process in almost every scientific domain. The problem involves finding the positions of a given pattern (usually of short length) in a reference stream of data (usually of large length). The matching can be an exact or as an approximate (inexact). Exact matching is to search for the pattern without allowing for mismatches (or insertions and deletions) of one or more characters in the pattern), while approximate matching is the opposite. For exact matching, several data structures that can be built in linear time and space are used and in practice nowadays. For approximate matching, the solutions proposed to solve this matching are non-linear and currently impractical. In this paper, we designed and implemented a structure that can be built in linear time and space ($O(n)$) and solves the approximate matching problem in $O(m + \frac {log_2n {(log_\Sigma n)} ^{k+1}}{k!} + occ)$ search costs, where $m$ is the length of the pattern, $n$ is the length of the reference, and $k$ is the number of tolerated mismatches (and insertion and deletions). 
\end{sciabstract}

\section*{Introduction}
Pattern matching is a fundamental problem in many scientific fields and their applications are tremendous and in practice unstoppably over the globe. Almost every aspect of our lives involves searching for data in a reference of a short size (small document or small database) or big size (DNA data, internet webpages, banking data, etc). The inputs are a text ($S$) of length $n$ over an alphabet of size $\Sigma$, pattern ($P$) of length $m$, and an integer number ($k$) of allowed errors (mismatch, insertion, or deletion). The outputs are the starting positions in $S$ of the sub-sequences that are at $k$ Hamming (or edit) distance with $P$. The simplest form of pattern matching, referred to as exact matching, is when the $k$ value is zero. This form was solved by several structures in optimal time and space (linear) \cite{hakak2019exact}. This includes mainly suffix trees \cite{weiner1973linear, mccreight1976space, ukkonen1995line}, suffix arrays \cite{abouelhoda2004replacing}, and FM-index \cite{ferragina2000opportunistic}. While approximate pattern matching in which the value of $k$ is one or more has not been solved optimally (in linear time and space). Approximate pattern matching is the focus of this paper. Several solutions proposed for approximate pattern matching\cite{canzar2015short, kucherov2016approximate}, but with impractical time and space. So, current solutions depend on structures that solve the exact matching followed by heuristic techniques to obtain results in practical time and space. Tools that are solving reads-to-genome alignment problems are examples of this approach \cite{alser2021technology,kucherov2019evolution}.

Given the larger constant factor of building suffix trees when compared to other linear structures such as suffix arrays and FM-index, the design of suffix tree structure is more flexible and dynamic to tackle string problems. This flexibility and dynamicity can be proven by looking at the number of problems that were solved so far by suffix tree rather than suffix array and FM index. The structures proposed in this paper are a continuation and improvement of the non-linear tree structure proposed as part of the PhD dissertation \cite{al2016novel} of the first author and was published in this article also \cite{al2015error}. The name of the structure is error tree ($ET$) and is built on the top of the suffix tree. In this paper, we present a linear ($O(n)$) design of $ET$ that can solve the approximate matching problem in $O(m + log_2n \frac {\Sigma^{k} log_\Sigma^{k}n}{k!} + occ$); noting that the number of strings that are at $k$ Hamming distance with a string of length $m$ is $O(\frac {\Sigma^{k-1} m^k}{k!})$.

\section*{Methods}
After building a suffix tree ($ST$) for the input data, building $ET$ can be in linear time and space for resolving the approximate pattern matching problem. Here we are describing the steps conceptually, where the technical and implementation details are provided in the Supplementary material. 

\subsection*{\textit{OSHR} tree structure}
The first and key step in building $ET$ in linear time was motivated by the following observation. Let's assume node $A$ has a suffix-link to node $B$, then the label (concatenation of all edges' labels) between node $A$ and each leaf node under $A$ \textbf{must} be presented between node $B$ and one of its leaf nodes (see Figure1). This means that any indexing (processing) of the suffixes under node $A$ can be applied implicitly at node $B$ without re-indexing (reprocessing) these same suffixes when indexing suffixes under node $B$. Therefore, the only suffixes that will need explicit processing (indexing) under node $B$ are the ones, if any, that were not presented under any of the $\Sigma$ nodes which have suffix-links to node $B$. 

This indexing schema requires that all nodes which have suffix-links to a node, let's say node $A$, must be indexed (processed) before indexing (processing) node $A$. In addition, all nodes in the suffix tree must be indexed (processed) in a recursive mode (postorder traversal). This urges the revealing and construction the following tree structure by reversing the suffix links in $ST$ so that: 
\begin{itemize}
\item Root node is the root of $ST$. 
\item Internal nodes are all internal nodes in $ST$ with at least one incoming suffix-link. 
\item Leaf nodes are all internal nodes in $ST$ with only outgoing suffix-link (no incoming suffix-links. 
\item There is a directed edge from node $a$ to node $b$ if $b$ has a suffix-link to node $a$.
\end{itemize}. 

In order to distinguish this tree structure from $ST$ tree and other tree structures, the name of this tree is $OSHR$ tree (the reason behind this acronym is provided in the Acknowledgment section). Leaf nodes in $ST$ are not included in the $OSHR$ tree as there is no outgoing nor incoming suffix-link from or to these nodes. Note that by the construction properties of $ST$ and suffix links, $OSHR$ tree will be a directed acyclic graph. Clearly, the space and time costs for building $OSHR$ tree are linear. The tree structure can be built implicitly (inside $ST$ tree) or explicitly (outside $ST$). 

\subsection*{$OT$ indexing}
The key building block of this indexing schema is the following two observations.

Firstly, which is the key one in construction the aforementioned index, is the following. If node $a$ has a suffix link to node $b$, then all the set of suffixes \textit{under} node $a$, denoted as subset ${A}$, must be a subset or equal to the set of the suffixes under node $b$ denoted as subset ${B}$. This indicates that if we assign index values to the suffixes in subset ${A}$, then these suffixes will be implicitly indexed in subset ${B}$ and we just need to assign new $OT$ index values to the indexes ${B-A}$. Note that this process will work recursively, in other words, if node $b$ has a $suffix\_link$ to node $c$, then we will just need to assign $OT$ index values to the set ${C-B}$ where ${C}$ is the set of suffixes under node $c$ and there will be no computation or indexing process associated with the set of ${C-A}$ as they are already covered in $OT$ index.    

Secondly, the structure of $ST$ includes the fact that an internal node, let's say node $x$, may have up to $O(\Sigma)$ nodes with suffix link linking to it. Now, in order to build $OT$ index correctly, we should start indexing all suffixes under each node with suffix links linking to node $x$ before indexing node $x$. This indicates a postorder traversal process, hence, we must construct a tree structure in order to perform this postorder traversal.

Therefore, indexing (processing) suffixes/strings under all nodes that have suffix-links to a node ($x$) and not re-indexing (re-processing) these suffixes under node $x$ through a postorder traversal of the $OSHR$ tree is defined as $OT$ indexing (processing, the reason for $OT$ acronym is provided in the Acknowledgment section). 

As a simple example from Figure 1, through postorder traversal of $OSHR$ tree,  node 26 was reached. Then, node 15 must be visited and an $OT$ index values let's say 3 and 4 for suffix number 6 ("AATTTAACTAAG\$") and suffix number 9 ("TTAACTAAG\$") will be assigned. Now, declare and assign at the visited node the variable $left\_OT\_index$ variable with value 3 and $right\_OT\_index$ variable with value 4. Next, node 21 will be visited and similarly suffix with number 11 ("AACTAAG\$") and 10 ("TAACTAAG\$") under this node will be indexed where $left\_OT\_index$ variable with value 5 and $right\_OT\_index$ variable with value 6 will be declared and assigned. Now, when node 26 will be visited, only suffix of "AAG\$" will be indexed with an $OT$ index value of 7, hence $left\_OT\_index$ variable with value 3 and $right\_OT\_index$ variable with value 7 will be declared and assigned to this node. This way we could index all suffixes under node 26 without an explicit index (process) for all of them. Continue the traversal recursively until the root node is reached. 

The following two sections present two indexing algorithms using $OT$ indexing toward resolving the approximate pattern matching problem. 

\begin{figure}
  \includegraphics[width=\linewidth]{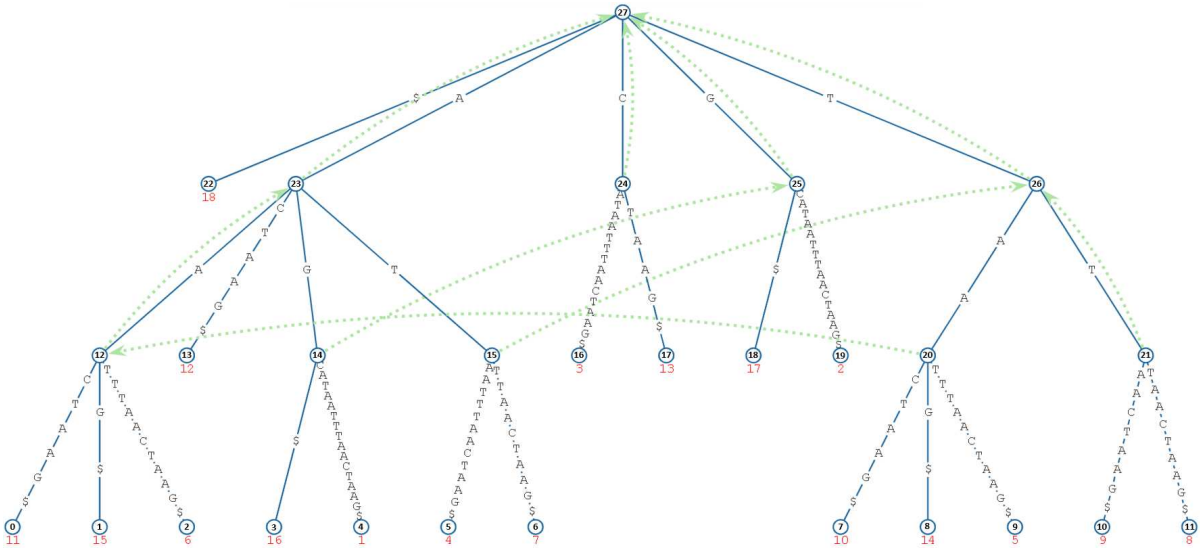}
  \caption{Suffix tree of string AGCCTAATTTAACTAAG\$ (drawn at https://hwv.dk/st/?AGCATAATTTAACTAAG\$). Suffix indexes start by 1 (not 0)}
  \label{fig:boat1}
\end{figure}

\subsection*{$OT$ indexing of base suffixes}
Note that an internal node, let's say node $x$, \textit{may} contain suffixes that have not existed under any of the nodes that have suffix link to node $x$. Let's denote such suffixes as base suffixes at node $x$. As examples from Figure 1: the base suffix under node 26 is the suffix "AAC\$", the base suffixes for node 23 are "TAATTTAACTAAG\$", "GCATAATTTAACTAAG\$", "ATTTAACTAAG\$", "AG\$", and "ACTAAG\$", and the base suffixes of node 12 are none. If a node has no incoming suffix links, then all suffixes under this node are base suffixes (node 20 as an example). The term "base" is selected to indicate that this is the first appearance of this suffix. The total number of base suffixes in all internal nodes will be always equal to $n$ (hence root node must have $left\_OT\_index$ of 0 and $right\_OT\_index$ of $n-1$, which was shown and proved in the implementations of this index). When the base suffixes at each internal node are computed, then $OT$ indexing of base suffixes under all internal nodes will cost linear time and space. Hence, all suffixes under all internal nodes will be indexed in linear time and space. 

Finding the base suffixes under all internal nodes can be computed trivially with $O(nh)$ time (where $h$ is the height of $ST$ tree) and $O(n)$ space by checking in all internal nodes whether each suffix that is under a node, let's say node $x$, is existed or not under any of the $\Sigma$ nodes that have suffix-links to node $x$. However, using both structures of $ST$ and $OSHR$ trees, base suffixes  of child (direct child) internal nodes of an internal node, nodes that link (back and forth) to an internal node and its child internal nodes, and several rules/tricks (that can be derived from python code snippet in Listing 1), all base suffixes for all internal nodes can be computed in linear time and space. Hence, $OT$ indexing of all base suffixes under all internal nodes can be computed in linear time and space ($O(n)$).

For the approximate matching problem and using $OT$ indexing of base suffixes, the following algorithm can be applied so that the search query cost can be $O(m + log_2n\frac {\Sigma^{k-1} log_\Sigma^{k}n}{k!} + occ$) but with $O(nh)$ indexing costs (time and space). For each base suffix at each node, find in the set of nodes in the path of the suffix in $ST$ starting from the root node, then assign an $OT$ index value for each node, let's say node $x$, in the set and record this value in a list (let this list be called $OT\_indexes$) already defined at node $x$. Let node $A$ has suffix-link to node $B$, note that once a base suffix is indexed at node $A$, then this indexing can be implicitly applied at node $B$. Therefore, as there are $n$ base suffixes and as the cost to find the nodes in the path for each suffix is $O(h)$, the total cost for this processing (and the total length of the final $OT$ index) will be $O(nh)$. When the indexing process is finished for all base suffixes at all internal nodes, each $OT\_indexes$ lists stored at different internal nodes must be sorted (given that they will be already sorted by default). This sorting is needed for lookup processes during the searching process. 

Now, the searching process for a given pattern and given $k$ value (let $k$ = 1 for a start) will proceed as follows. Walk the pattern in $ST$. If a mismatch occurs at an edge, continue walking as an exact match until the end of the pattern, let the last node of this walk is $x$, then output the occurrences which are indexes of all leaf nodes under node $x$. If the walk reaches an internal node, let's say node $x$ with node's depth $d$, without any mismatch, then lookup for $Left\_OT\_index$ and $Right\_OT\_index$ of node $x$ in $OT\_indexes$ list stored at end node of $\sigma p_{d+1}p_{d+2}....p_m$ where $\sigma$ is each value in the alphabet and $p_x$ is the character at position $x$ in the pattern. If there is at least one value in between the found positions of $Left\_OT\_index$ and $Right\_OT\_index$ in $OT\_indexes$, this means the string $\sigma p_{d+1}p_{d+2}....p_m$ was indexed under node $x$ and must be existed under node $x$, so walk the string $\sigma p_{d+1}p_{d+2}....p_m$ from node $x$ to find the occurrences of this string. If not, this means string $\sigma p_{d+1}p_{d+2}....p_m$ does not exist under node $x$ as it was not indexed. 

To find the occurrences of $K=2$, perform the same process above accordingly on the paths found for $K=1$. The same for $k\geq3$. In case $ST$ was not balanced, then treat the heaviest path during the search process as an edge. The cost for this search process will be $O(m + log_2n\frac {\Sigma^{k-1} log_\Sigma^{k}n}{k!} + occ$). As the cost of this index is non-linear, it can be efficient when the input text is small enough. In the following section, we will describe a linear $OT$ index with a minor increase in the search process cost.

\subsection*{OT indexing of base paths}
Before all, let's define an internal that at least one of its children is an internal node as a "grand node" and the internal node that all of its children are leaf nodes as a "non-grand node". The motivations for this $OT$ indexing are the following two observations. Firstly, the main complexity in the searching process for a pattern is caused by the branching caused by internal nodes. Secondly, if the search process reaches an internal node with depth $d$ where all of the nodes' children are leaf nodes (non-grand nodes), then the cost for finding approximate matching if mismatch positions would be positions $d$ to $d + k$ will cost $O(\Sigma)$ time (approximate matching each label of leaf node will cost $O(1)$ time so the total cost will be $O(\Sigma)$, detailed explanation of this cost is provided in the Supplementary Material). Hence, there is no need to index labels of leaf nodes under non-grand nodes or even the leaf nodes under the current node in the search process (these labels are indexed in the $OT$ indexing version which indexes the base suffixes), This way, indexing the paths between each internal node in the $ST$ and its descendant internal nodes will be sufficient and efficient to handle the approximate matching problem as explained in short. 

Finding and indexing all paths between each internal node and its descendant internal nodes cost $O(nh)$ using a trivial algorithm. However, due to the structure of $ST$, many paths are redundant throughout $ST$ structure. As an example, in the small $ST$ in Figure 1, the path between the root node and node 25 already exists between node 23 and node 14. So indexing the later path through $OT$ indexing will implicitly index the former one. So, paths that are needed to be explicitly indexed are \textit{base paths} which are the paths that \textit{firstly} appeared between two internal nodes and never appeared between any other two internal nodes. 

Theoretically, the distinct labels of all base paths \textit{must} eventually equal labels of paths between the root node and all descendant internal nodes. This is indeed must be true as the paths between the root node and all descendant internal nodes are the last appearance (or extent) of any base path in $ST$ through suffix-links(this also was proved by implementations and shown). Note that any path between the root node and any internal node can have up to $O(\Sigma)$ paths from the other two nodes (through suffix-links). As these  $O(\Sigma)$ paths must be originated from a base path, then the total number of base paths and their redundancies are no more than $O(\Sigma n)$. This means that indexing only base paths will cost $O(\Sigma n)$ time and space. Therefore, through $OT$ indexing of base paths, all paths between all internal nodes and their descendant internal nodes will cost $O(n)$ time and space. Building $OT$ index for base paths requires post-order traversal of $OSHR$ tree where at each visited grand node, only a selected set of internal nodes that are descendants to the visited node (descendants under $ST$ structure) are indexed. The node in the set forms a base path with the visited node where the selection process is based on three rules provided in Supplementary Material. 

During the searching process using $ST$ for let's say $k=1$, if the search reaches node $x$ at depth $d$, finding the deepest matching internal node, if any, will cost $O(\Sigma log_2nlog_\Sigma n)$ ($O(log_\Sigma n)$ is the expected number of nodes in the path in $ST$ of suffix $d+ 1$ of the pattern, $log_2n$ is the time needed to lookup in each aforementioned node for $left\_OT\_index$ and $right\_OT\_index$ values stored at node $x$ for each possible $\sigma$ in $\Sigma$). Once the deepest matching internal node (let's say node $m$) is reached, proceed as the following. If no more matching is needed then output the suffix indexes under node $m$. If more matching is needed and node $m$ is non-grand, complete the matching with the label of each leaf node ($O(\Sigma)$ time). If node $m$ is a grand node, complete the matching with the label of each leaf node where each will cost $O(1)$ and with the label of the edge between node $m$ and each of its child internal nodes (where each will cost $O(1)$), so the total cost will be $O(\Sigma)$ time. Note that there is no need to perform matching with any path under any child internal node of node $m$ as node $m$ was computed as the deepest possible matching internal node and the matching occurrence \textit{must} be inbetween node $m$ and one of its child internal nodes. If the matching occurrence is under a child internal node, let's say node $c$, of node $m$, then the deepest matching internal node would be node $c$ not node $m$. A full description of the search process is provided in Supplementary Material. Hence, the total cost for searching for Hamming distance of $K$ with a pattern of length $m$ will be $O(m + \frac {log_2n {(log_\Sigma n)} ^{k+1}}{k!} + occ)$. Edit distance can be handled accordingly.

\section*{Results}
The structures and both $OT$ indexes are implemented and provided at \url{https://github.com/aalokaily/}.  
For testing the implementation of $ET$, we used ten genomes, listed in Table 1, ranging in size from 50KB to 100MB. The time cost for each step involved in the linear building $ET$ structure is listed in Table 1. Table2 shows the time needed to build an error tree for each genome. Note that building of $OT$ index took time (and space) close to building $ST$. This applies also to the step of preprocessing procedures.

\textbf{Acknowledgment}
The name $OSHR$ tree stands for Okaily-Sheehy-Huang-Rajasekaran which is the last name of the first author and the last names of his PhD committee in the University of Connecticut, Department of Computer Science, where the initial version of the error tree structure was a chapter in the PhD dissertation. The committee members were: Chun-Hsi Huang
(Major Advisor), Sanguthevar Rajasekaran, and Don Sheehy. This meant to tribute to them and appreciate their kind,
influential, and professional teaching and supervision. The name $OT$ stands for the last names of the authors of this work.

\renewcommand{\refname}{References and Notes}

\bibliography{scifile}

\bibliographystyle{Science}


\section*{Supplementary materials}
In a general analysis of suffix trees, there are main challenges that are easily faced and needed to be carefully addressed in order to resolve string problems, such as approximate matching problem, more efficiently or optimally. Firstly, under different internal nodes there are similar suffixes; how can we track these same suffixes across different nodes so that if a process is performed on a single suffix then we can apply or just link the outcomes of this process to the same suffix in different node/s and save the costs of performing the same process again and again. Secondly, the structure under an internal node is symmetric partially or fully to the structure under other internal nodes (we mean by "partially symmetric" a subtree under an internal node is symmetric with the subtree under another internal node/s); how can we trace and embed this interconnectivity across different internal nodes.    

The possible answer to the first challenge is to build a global index for each $O(n)$ suffixes, then record the index value of each suffix for each suffix under each internal node. Away from the $O(nh)$ cost of this indexing schema, the index values that will be recorded in some internal node will be distributed randomly. This will lead to a costly computation when the index values of two different nodes need to be compared or intersected for some purposes. In addition, given that this indexing schema will be useful when two internal nodes are fully symmetric, it will not be highly useful in case two internal nodes are partially symmetric or asymmetric (which are the common cases). So, the indexing schema can not be arbitrary and has to follow some structure and take into account (and take advantage) of the interconnectivity among different internal nodes, but how to find or create this structure and from which suffix/node should I start the indexing schema and in which order. These questions are the challenges and motivation behind this paper and their answers are the goals of this paper.

\section{Suffix tree construction}
Algorithms proposed in this paper were implemented using python language (python3). For building $ST$, we used a python package \url{https://pypi.org/project/suffix-trees/} in which $index$, $depth$, $parent$, and $suffix\_link$ are implemented for each node in the tree and with removing the Snode \_\_slots\_\_ line to allow setting attributes more freely. $Depth$ attribute at each node which is equal to the lengths of all edges from root to the node, $Suffix\_index$ attribute at each node which is equal to the index in $S$ of the starting character of string extracted from root to that node,  $parent$ attribute at each node that stores the memory location of the parent node, and $Suffix\_links$ (which must be originally constructed in $ST$) should be preserved. 

\section{Preprocessing procedures}
Next, the following preprocessing steps were computed by postorder traversing of $ST$: 
\begin{itemize}
    \item Assign serialized keys to leaf nodes from left to right. 
    \item Construct $OSHR$ tree as shown in the code snapshot below. This code is applied within postorder traversal of $ST$. Note that an internal node that has no incoming suffix link is referred as $OSHR$ leaf node whereas the internal node that does is referred as $OSHR$ internal node. These concepts were coded and used in this work. 
    
    \begin{lstlisting}[language=Python, caption=procedures for building \textit{OSHR} tree]
    # construct implicit OSHR tree
					if current_node._suffix_link is not None and current_node != tree.root:
						temp = current_node._suffix_link
						if not hasattr(temp, "nodes_link_to_me"):
							setattr(temp, "nodes_link_to_me", [])
						temp.nodes_link_to_me.append(current_node)
    
    \end{lstlisting}
    
    \item Set attributes of \textit{key\_of\_leftmost\_leaf\_node} and \textit{key\_of\_rightmost\_leaf\_node} to each internal node where the value of \textit{key\_of\_leftmost\_leaf\_node} equals the key of leftmost leaf node under that internal node (\textit{key\_of\_rightmost\_leaf\_node} likewise). 
    
    \item Create two auxiliary lists. The first list is to store the suffixes indexes of leaf nodes from left to right, denoted hereafter and in the code snapshots as "left\_to\_right\_suffix\_indexes\_list". The second list maps suffix indexes of leaf nodes to their memory locations as an example, if the suffix index stored at the leaf node is $x$, then store at position $x$ of the list the memory location of the leaf node (this list helps to speed access finding and accessing the memory location of leaf nodes as needed, and is denoted hereafter and in the code snapshots as "leaf\_suffix\_index\_to\_leaf\_memory\_list"). The size of each list is linear (equals to the number of leaf nodes).

    The following preprocessing procedures are needed for OT indexing of base paths only in additions to the above ones. The last two procedures need the $OSHR$ tree structure to be already built, so after performing the above procedures in a postorder traversal (in which the $OSHR$ tree is built), traverse the $ST$ for a second time and perform the following:
    
    \item Let node $A$ be a parent node of an internal node $B$ and node $A$ has suffix link to node $C$ and node $B$ has suffix link to node $D$. If node $C$ is not the direct parent of node $D$, then for each node inbetween nodes $C$ and $D$: mark the node as "inbetween\_node" and record/store node $B$ to be reference node for this node. These preprocessing procedures will be used in later phases as will be explained.  
    
    \item Store $OSHR$ leaf nodes from left to right and record at each internal node two values \textit{index\_of\_leftmost\_OSHR\_leaf} and \textit{index\_of\_rightmost\_OSHR\_leaf} (the index values are the indexes in the created list).
    
    \item Store $OSHR$ internal nodes from left to right and record at each internal node two values \textit{index\_of\_leftmost\_OSHR\_internal} and \textit{index\_of\_rightmost\_OSHR\_internal} (the index values are the indexes in the created list).

\end{itemize}

\section{OT indexing}

For running $OT$ indexing, an iterative postorder traversal of the $OSHR$ tree must be computed. Once a processing/indexing is performed under a visited node, this processing/indexing can be applied implicitly to all reaming nodes in the traversal. 

\section{Finding base suffixes}
Base suffix is the first appearance of a suffix under an internal node, let's say $a$ (where suffix label starts from node $a$ to the leaf node of the suffix) and never was appeared under any node that has suffix link to node $a$. This is the reason for denoting this suffix as a base suffix. Suffixes suffixes that are not base suffixes are considered to be and referred to as extent/extension suffixes. The last extent suffix must start from the root node. The total number of base suffixes must be equal exactly to $n$. Once a base suffix appeared under node $a$, it must be appeared under the node that node $a$ has suffix link to, and continually, until the root node is reached. Therefore, once a base suffix is processed/indexed, this processing/indexing can be applied implicitly to all (expectedly ($log_\Sigma n$)) extent suffixes.

Finding base suffixes can be computed naively with time cost of $O(nh)$ and space cost of $n$. This can be performed by traversing $OSHR$ or $ST$ tree and checking (can be in constant time) if each suffix under the visited node was recorded ever as a base suffix. If not, it's a base suffix and record it under the visited node. The time cost will be $O(nh)$ as the leaf nodes under each internal node in the $OSHR$ tree should be processed. However, as the upper bound of the number of base suffixes is $O(n)$, then an algorithm that could find each base suffix in constant time will need linear costs. The following two algorithms provide a linear cost for finding base suffixes (the first one is faster).

\begin{lstlisting}[language=Python, caption=Algorithm for finding base suffixes]
def phase_1_for_OT_indexing_for_base_suffixes(tree):	
		# find base suffixes
		stack.append(tree.root)
		children_stack.append((list(tree.root.transition_links[x] for x in sorted(tree.root.transition_links.keys(), reverse=True))))		
		cost = 0
		
		while stack:
			current_node = stack[-1]				  
			if len(children_stack[-1]) > 0:
				last_node_under_top_node_in_stack = children_stack[-1][-1]
				stack.append(last_node_under_top_node_in_stack)	 
				
				children_stack[-1].pop()					
				children_stack.append((list(last_node_under_top_node_in_stack.transition_links[x] for x in sorted(last_node_under_top_node_in_stack.transition_links.keys(), reverse=True))))										   
			else:
				stack.pop()
				children_stack.pop()
				
				# alongside processing
				#print (current_node.key)
				setattr(current_node, "OT_indexes", []) # to be used in next phase 
				for child_node in current_node.transition_links.values(): # current_node.transition_links.values() contains child nodes of current_node as part of ST structure
					if child_node.is_leaf():
						if child_node.idx + 1 < tree.number_leaf_nodes:
							leaf_node_of_next_suffix_index = tree.leaf_suffix_index_to_leaf_memory_list[child_node.idx + 1]									   
							if leaf_node_of_next_suffix_index.parent != current_node._suffix_link:
								temp = leaf_node_of_next_suffix_index.parent
								if current_node == tree.root:   # then we must include the tree.root in the process
									while True:
										if hasattr(temp, "nodes_link_to_me"):    # this condition made to skip the OSHR leaf nodes as these nodes "already" collect uncovered suffixes by retrieving all suffix indexes below them. So there is no need 
																			    # to add (in fact duplicate) them to these nodes. We may avoid this condition and speed up the process a bit by create a link between an OSHR internal node and its first
																			    # ancestor OSHR internal node, and use this link in the loop; but at the cost of extra space for saving these links for sure.
											temp.base_suffixes.append(leaf_node_of_next_suffix_index.idx + temp.depth)
										if temp == tree.root:	
											break
										else:
											temp = temp.parent
										cost += 1
								else:
									end_node = current_node._suffix_link
									while temp != end_node:
										if  hasattr(temp, "nodes_link_to_me"):
											temp.base_suffixes.append(leaf_node_of_next_suffix_index.idx + temp.depth)
										temp = temp.parent
										cost += 1
									
				if not current_node.is_leaf():
					#find and mark inbetween top base node and assign the reference nodes for this node (which are as coded below)
					top_node = current_node.parent._suffix_link 
					bottom_node = current_node._suffix_link.parent
					if bottom_node != top_node:
						n = bottom_node
						while n  != top_node:
							if hasattr(n, "nodes_link_to_me"):# if node is an OSHR leave node (has no nodes_link_to_me attribute) then this case is already handled 
								for leaf_node_index in tree.left_to_right_leaf_nodes_list[current_node.index_of_leftmost_leaf:current_node.index_of_rightmost_leaf + 1]:
									n.base_suffixes.append(leaf_node_index + 1 + n.depth)
									cost += 1
							n = n.parent
							
					if  not hasattr(current_node, "nodes_link_to_me"):
						for leaf_node_index in tree.left_to_right_leaf_nodes_list[current_node.index_of_leftmost_leaf:current_node.index_of_rightmost_leaf+1]:
							current_node.base_suffixes.append(leaf_node_index  + current_node.depth)
							
						cost += current_node.index_of_rightmost_leaf -  current_node.index_of_leftmost_leaf + 1
		
		# compute the case for suffix 0 as there is previous index for inddex 0 
		leaf_node_of_suffix_index_zero = tree.leaf_suffix_index_to_leaf_memory_list[0]									   
		if leaf_node_of_suffix_index_zero.parent != current_node._suffix_link:
			temp = leaf_node_of_suffix_index_zero
			while temp != tree.root:
				temp = temp.parent
				if  hasattr(temp, "nodes_link_to_me"):
					temp.base_suffixes.append(0 + temp.depth)
					
		# this a special cases and for the root only. The suffix-link of child internal node of a root usually link to the root. In case not, 
		# then the node that the child internal node link to must be bottom-node for the root node.
		current_node = tree.root
		for node in current_node.transition_links.values():
			if node.is_leaf():
				if node.idx + 1 < tree.number_leaf_nodes:
					leaf_node_of_next_suffix_index = tree.leaf_suffix_index_to_leaf_memory_list[node.idx + 1]
					if leaf_node_of_next_suffix_index.parent == tree.root:
						temp.base_suffixes.append(leaf_node_of_next_suffix_index.idx + temp.depth)
			
			else:
				tt = node._suffix_link
				if tt != tree.root and tt.parent != tree.root:
					for leaf_node_index in tree.left_to_right_leaf_nodes_list[tt.index_of_leftmost_leaf:tt.index_of_rightmost_leaf+1]:
						tree.root.base_suffixes.append(leaf_node_index)
						
		print ("Finding base suffixes took", cost)	

	stack = []
	children_stack = []
	start = time.time()
	phase_1_for_OT_indexing_for_base_suffixes(tree)

\end{lstlisting}

\begin{lstlisting}[language=Python, caption=Algorithm for finding base suffixes]
def phase_1_for_OT_indexing_for_base_suffixes_not_used(tree):	
		# this is the first but slower algorithm for finding base suffixes 
		stack.append(tree.root)
		children_stack.append((list(tree.root.transition_links[x] for x in sorted(tree.root.transition_links.keys(), reverse=True))))		
		count = 0
		while stack:
			current_node = stack[-1]				  
			if len(children_stack[-1]) > 0:
				last_node_under_top_node_in_stack = children_stack[-1][-1]
				stack.append(last_node_under_top_node_in_stack)	 
				
				children_stack[-1].pop()					
				children_stack.append((list(last_node_under_top_node_in_stack.transition_links[x] for x in sorted(last_node_under_top_node_in_stack.transition_links.keys(), reverse=True))))										   
			else:
				stack.pop()
				children_stack.pop()
				
				# alongside processings
				#print (current_node.key)
				temp_base_suffixes_list = []
				if not current_node.is_leaf():						   
					if hasattr(current_node, "nodes_link_to_me"):  
					# if node has nodes_link_to_me attribute then it's an internal node in OSHR tree 
					# where the nodes that link to it are stored in this attribute					  
						# collect data from nodes_link_to_me attribute
						temp_base_suffixes_list = []
						nodes_linked_to_me = current_node.nodes_link_to_me
						s = 0								
						for node_linked_to_me in nodes_linked_to_me:
							s += node_linked_to_me.index_of_rightmost_leaf - node_linked_to_me.index_of_leftmost_leaf + 1
							count += 1
						# now the value of s is the sum of all suffix indexes under all nodes that link to current_node in the traversal
						if s != current_node.index_of_rightmost_leaf - current_node.index_of_leftmost_leaf + 1:						 
							# if the above condition is false, then there will be no uncovered suffixes at all to search for, as the sum of suffixes under nodes link to 
							# current_node are equal to the number of suffix indexes under current_node
							for child_node in current_node.transition_links.values(): # current_node.transition_links.values() contains child nodes of current_node as part of ST structure
								if child_node.is_leaf():
									# The following lines check whether the the previous suffix index of the suffix of the child leaf node was covered under any of the nodes that link to current_node. 
									# If not covered then add it to the base_suffixes list of current_node
									count += 1
									leaf_node_of_previous_suffix_index = tree.leaf_suffix_index_to_leaf_memory_list[child_node.idx - 1]									   
									if leaf_node_of_previous_suffix_index.parent._suffix_link != current_node:											  
										f = True
										for node_linked_to_me in nodes_linked_to_me:
											count += 1
											if leaf_node_of_previous_suffix_index.key in range(node_linked_to_me.index_of_leftmost_leaf, node_linked_to_me.index_of_rightmost_leaf + 1):
												f = False
												break
										if f:
											count += 1
											temp_base_suffixes_list.append(child_node.idx)									   
								
								else: 
									if hasattr(child_node, "nodes_link_to_me"): # This means child_node is an internal node in OSHR tree
										a = 0
										for node_links_to_child_node in child_node.nodes_link_to_me:
											# The following lines code a tricky process which compute that if the condition is true, then all suffix indexes under node_links_to_child_node must be 
											# in the uncovered suffixes under current node 
											count += 1
											if node_links_to_child_node.parent._suffix_link != current_node:										 
												for suffix_idx in tree. left_to_right_suffix_indexes_list[node_links_to_child_node.index_of_leftmost_leaf:node_links_to_child_node.index_of_rightmost_leaf + 1]:
													temp_base_suffixes_list.append(suffix_idx + 1)	
													count += 1
											else:
												a += 1
												count += 1
										
										if hasattr(child_node, "base_suffixes"):
											if a == len(nodes_linked_to_me):
												temp_base_suffixes_list += child_node.base_suffixes
												count += len(child_node.base_suffixes)
												# the above condition cover a special and common case in order to speed up the processing and avoid the computation in else statement below
											else:
												# The following lines check whether the the previous suffix index of the suffixes in child_node.base_suffixes list was covered under any of the nodes that 
												# link to current_node. If not covered then add it to the base_suffixes list of current_node
												for suffix_idx in child_node.base_suffixes:
													count += 1
													f = True
													key_of_prev_idx_node = tree.leaf_suffix_index_to_leaf_memory_list[suffix_idx - 1].key
													for node_linked_to_me in nodes_linked_to_me:
														count += 1
														if key_of_prev_idx_node in range(node_linked_to_me.index_of_leftmost_leaf, node_linked_to_me.index_of_rightmost_leaf + 1):
															f = False
															break
													if f:
														temp_base_suffixes_list.append(suffix_idx)	
														count += 1
									else:
										# means child_node is a leaf node in OSHR tree (no suffix_link is linking to it), as so, check whether the the previous suffix index of the suffixes in 
										# child_node.base_suffixes list was covered under any of the nodes that link to current_node. If not covered then add it to the base_suffixes list of current_node
										for suffix_idx in child_node.base_suffixes:
											count += 1
											leaf_node_of_previous_suffix_index = tree.leaf_suffix_index_to_leaf_memory_list[suffix_idx - 1]											   
											if leaf_node_of_previous_suffix_index.parent._suffix_link != current_node:
												f = True
												for node_linked_to_me in nodes_linked_to_me:
													count += 1
													if leaf_node_of_previous_suffix_index.key in range(node_linked_to_me.index_of_leftmost_leaf, node_linked_to_me.index_of_rightmost_leaf + 1):
														f = False
														break
												if f:
													temp_base_suffixes_list.append(suffix_idx)	
													count += 1
							setattr(current_node, "base_suffixes", temp_base_suffixes_list)
						else:
							setattr(current_node, "base_suffixes", [])
							count += 1
					else:	# means current_node is a leaf node in OSHR tree (no suffix_link is linking to it), as so, just add all suffix indexes under current_node to the base_suffixes list of the node (itself) 
						temp_base_suffixes_list += tree. left_to_right_suffix_indexes_list[current_node.index_of_leftmost_leaf:current_node.index_of_rightmost_leaf+1]
						setattr(current_node, "base_suffixes", temp_base_suffixes_list)  
						count += 1
	
						
		
				#if hasattr(current_node, "base_suffixes"):
				#	print (current_node.key, current_node.base_suffixes, current_node.is_leaf())
					
		print ("Finding uncovered suffixes took", count)	
	stack = []
	children_stack = []
	start = time.time()
	
\end{lstlisting}

\section{Finding base paths}
Base path is a path between two nodes (top and bottom nodes) where the label between both nodes never has appeared between any other two internal nodes. There must be an extension path from a base path induced throughout suffix links of top and bottom nodes. The total number of the distinct labels of all base paths is equal exactly to the number of internal nodes (except the root node) and must match the labels between root node and each internal node in the tree. Similar to base suffixes, once a base path is processed/indexed, all extent/extension paths are processed/indexed implicitly.

Finding base paths naively can cost $O(nh)$ time and $n$ space (explanation is omitted as it's similar to the base suffix case). However, using the following rules, the process can be linear. Traverse $ST$, if visited node is an $OSHR$ leaf node (no incoming suffix links) or $OSHR$ internal node, then all paths between the visited node and its descendant $OSHR$ leaf nodes are base paths. If the visited node is marked as inbetween node and is an $OSHR$ leaf node, then all paths between the visited node and its descendant $OSHR$ internal nodes are base paths. If is an $OSHR$ internal node, then all paths between the visited node and the nodes that have incoming suffix links from the descendants internal nodes under each reference node (reference nodes of visited node). The following code snapshot presents a linear algorithm for finding base suffixes. There are special cases costing constant time and space described in the code snapshot.

\begin{lstlisting}[language=Python, caption=Algorithm for finding base paths]
	def phase_2_for_OT_indexing_for_base_paths(tree, k):
		# find base paths, record base and bottom nodes, and create OT index
		global text
		stack.append(tree.root)
		children_stack.append((list(tree.root.nodes_link_to_me)))
		key_stack.append(0)
		
		while stack:
			current_node = stack[-1]	
			# check if OSHR[current_node.key] is empty, then remove it from stack
			if len(children_stack[-1]) > 0:
				last_node_under_top_node_in_stack = children_stack[-1][-1]
				stack.append(last_node_under_top_node_in_stack)		
				children_stack[-1].pop()
				if hasattr(last_node_under_top_node_in_stack, "nodes_link_to_me"):
					children_stack.append((list(last_node_under_top_node_in_stack.nodes_link_to_me)))
				else:
					children_stack.append([])

				key_stack.append(tree.keys_counter + 1)	 
				   
			else:
				# collect bottom-base nodes that are OSHR leaf nodes 	
				OSHR_leaf_nodes = []
				if hasattr(current_node, "index_of_leftmost_OSHR_leaf"):
					OSHR_leaf_nodes = tree.OSHR_leaf_nodes_left_to_right_list[current_node.index_of_leftmost_OSHR_leaf:current_node.index_of_rightmost_OSHR_leaf + 1]
				
				# collect bottom-base nodes collected from refrerence nodes if current_node is inbetween_top_base_node 
				inbetween_bottom_base_node_dict = defaultdict()  # this dict will be used to ditinict nodes under tow difference refeence nodes that are linking to the same node under current_node					
				inbetween_bottom_base_node_list = []
				if hasattr(current_node, "inbetween_top_base_node"):
					if hasattr(current_node, "nodes_link_to_me"):
						inbetween_bottom_base_node_dict = defaultdict()  # this dict will be used to ditinict nodes under tow difference refeence nodes that are linking to the same node under current_node					
						for referennce_node in  current_node.inbetween_top_base_node:
							inbetween_bottom_base_node_dict[referennce_node._suffix_link.key] = referennce_node._suffix_link
							if hasattr(referennce_node, "index_of_leftmost_OSHR_leaf"):
								for node in tree.OSHR_leaf_nodes_left_to_right_list[referennce_node.index_of_leftmost_OSHR_leaf:referennce_node.index_of_rightmost_OSHR_leaf + 1]:
									inbetween_bottom_base_node_dict[node._suffix_link.key] = node._suffix_link
							if hasattr(referennce_node, "index_of_leftmost_OSHR_internal"):
								for node in  tree.OSHR_internal_nodes_left_to_right_list[referennce_node.index_of_leftmost_OSHR_internal:referennce_node.index_of_rightmost_OSHR_internal + 1]:
									inbetween_bottom_base_node_dict[node._suffix_link.key] = node._suffix_link
					else:
						if hasattr(current_node, "index_of_leftmost_OSHR_internal"):
							inbetween_bottom_base_node_list  = tree.OSHR_internal_nodes_left_to_right_list[current_node.index_of_leftmost_OSHR_internal:current_node.index_of_rightmost_OSHR_internal + 1]
					
				# the following 6 lines cover a special case and for the root only. The suffix-link of child internal node of a root usually link to the root. In case not, 
				# then the node that the child internal node link to must be bottom-node for the root node.
				root_bottom_nodes = []
				if current_node == tree.root:
					for node in current_node.transition_links.values():
						if not node.is_leaf():
							if node._suffix_link != tree.root:
								root_bottom_nodes.append(node._suffix_link)


						
				for bottom_base_node in list(inbetween_bottom_base_node_dict.values()) + OSHR_leaf_nodes + inbetween_bottom_base_node_list + root_bottom_nodes:
					tree.keys_counter += 1
					mapping_guided_suffix = tree.left_to_right_leaf_nodes_list[bottom_base_node.index_of_leftmost_leaf]
					suffix_starting_from_current_node = mapping_guided_suffix + current_node.depth + k
					if suffix_starting_from_current_node < tree.number_leaf_nodes:
						index_key_of_suffix_starting_from_current_node_in_ST = tree.leaf_suffix_index_to_leaf_memory_list[suffix_starting_from_current_node].key
														
					if bottom_base_node.depth - current_node.depth - k not in tree.temp_dict:
						tree.temp_dict[bottom_base_node.depth - current_node.depth - k] = []									
					tree.temp_dict[bottom_base_node.depth - current_node.depth - k].append((index_key_of_suffix_starting_from_current_node_in_ST, tree.keys_counter, text[bottom_base_node.idx + current_node.depth:bottom_base_node.idx + current_node.depth + k]))
					tree.OT_index[tree.keys_counter] = (mapping_guided_suffix, current_node.depth, bottom_base_node.depth)
					

				
					
					
				if hasattr(current_node, "leftmost_OT_index"):
					current_node.leftmost_OT_index = key_stack[-1]
					current_node.rightmost_OT_index = tree.keys_counter
				else:
					setattr(current_node, "leftmost_OT_index", defaultdict())
					setattr(current_node, "rightmost_OT_index", defaultdict())
					
					current_node.leftmost_OT_index = key_stack[-1]
					current_node.rightmost_OT_index = tree.keys_counter
				
					
				key_stack.pop()
				stack.pop()
				children_stack.pop()
		
		
	stack = []
	children_stack = []
	
	tree.keys_counter = defaultdict(int)
	key_stack = defaultdict(int)		
	setattr(tree, "OT_index", defaultdict())
	
	tree.keys_counter = -1
	key_stack = []
	
		
	start = time.time()
	phase_2_for_OT_indexing_for_base_paths(tree, K)
\end{lstlisting}

\section{Resolving approximate pattern matching using OT indexing}
Note that $OT$ index can be useful for different string processing problems not only for the approximate pattern matching. For approximate pattern matching problem, several algorithms can be applied to resolve the problem more efficiently/optimally. 

\begin{itemize}
 \item Indexing base suffixes with $O(nlog_\Sigma n)$ construction cost (assuming $h$ value is on average $O(log_\Sigma n)$) and $O(m + \frac {log_2n {(log_\Sigma n)} ^{k}}{k!} + occ)$ searching cost. There are two methods for indexing base suffixes. 
 
\item Indexing base uncle suffixes with near-linear construction cost and $O(m + \frac {log_2n {(log_\Sigma n)} ^{k}}{k!} + occ)$ searching cost. Further descriptions of this indexing will be stated in short below.

\item Indexing base paths with linear construction cost and $O(m + \frac {log_2n {(log_\Sigma n)} ^{k+1}}{k!} + occ)$ searching cost.
\end{itemize}

The selection of algorithm is depending on the size of input data, expected speed of search process, expected number of search queries, and/or capacity of memory. Full python3 programs for the above four algorithms are provided at \url{https://github.com/aalokaily/}. 

\subsection{\textit{OT} indexing using base suffixes}
The first step is to find base suffixes and record them at the internal node where they appeared first (recording them as a number where suffix is started from the visited node (not from the root node). A linear algorithm is given in Section 4 for this step. Next, we need to map each base suffix under each internal node to its last extent suffix (the one that starts from the root node).   

Initialize an $OT\_index\_counter$ variable. Next, traverse the \textit{OSHR} tree in postorder and for each base suffix recorded at the visited node do the following:
\begin{itemize}
 \item Find the set of nodes in the path of base suffix in $ST$ starting from root. For each node increment $OT\_index\_counter$ variable by one and store the value in a list stored already at the node (let this list be denoted as $OT\_indexes$).  
 \item Map using a list or dictionary, let be denoted as $OT\_index$, the value of $OT\_index\_counter$ with the index of base suffix. This is needed in the search process. 
 
 Once all base suffixes recorded at the visited node have been $OT$ indexed, assign at the visited node two variables which are $left\_OT\_index$ and $right\_OT\_index$ where the first variable stores the value of $OT\_index\_counter$ before indexing any base suffixes at the visited node (stored in a recursive stack) and the later variable stores the last value of $OT\_index\_counter$. 
\end{itemize}

\begin{lstlisting}[language=Python, caption= OT indexing function using base suffixes (trivial/naive algorithm)]
def phase_2_for_OT_indexing_for_base_paths(tree, k):
		global text
		stack.append(tree.root)
		children_stack.append((list(tree.root.nodes_link_to_me)))
		key_stack.append(0)
		
			
		while stack:
			current_node = stack[-1]	
			# check if OSHR[current_node.key] is empty, then remove it from stack
			if len(children_stack[-1]) > 0:
				last_node_under_top_node_in_stack = children_stack[-1][-1]
				stack.append(last_node_under_top_node_in_stack)		
				children_stack[-1].pop()
				if hasattr(last_node_under_top_node_in_stack, "nodes_link_to_me"):
					children_stack.append((list(last_node_under_top_node_in_stack.nodes_link_to_me)))
				else:
					children_stack.append([])

				key_stack.append(tree.keys_counter + 1)	 
				   
			else:
								
				for base_suffix in current_node.base_suffixes:
					suffix_idx = base_suffix - current_node.depth
					node = tree.leaf_suffix_index_to_leaf_memory_list[base_suffix]
					transition_letter =  text[node.idx + current_node.depth:node.idx + current_node.depth + k]
					
					while node.depth > 0:
						tree.keys_counter += 1
						mapping_guided_suffix = base_suffix
						OT_indx = tree.keys_counter
						
						node.OT_indexes.append((OT_indx, transition_letter))
						node = node.parent	
						
				if hasattr(current_node, "left_OT_index"):
					current_node.left_OT_index = key_stack[-1]
					current_node.right_OT_index = tree.keys_counter
				else:
					setattr(current_node, "left_OT_index", defaultdict())
					setattr(current_node, "right_OT_index", defaultdict())
					
					current_node.left_OT_index = key_stack[-1]
					current_node.right_OT_index = tree.keys_counter	
				key_stack.pop()
				stack.pop()
				children_stack.pop()
				
\end{lstlisting}				
The above algorithm is a trivial version for indexing base suffixes as it indexes every nodes in the path of every base suffix, while in fact these nodes may intersect in the paths of other base suffixes. Detection of these intersections and avoiding their computations can be achieved by indexing the label of \textit{tails} of base suffixes (tail of base suffix is the edge between parent of leaf node of base suffix to the leaf node of base suffix) not all label of base suffixes (path between visited node to leaf node of base suffix). Let the currently visited node is node $a$ and suffix $x$ is a base suffixes under node $a$. Now, the path between node $a$ and the leaf node of base suffix $x$ (which must have a suffix index value of $x$ + depth($a$)) may contain several nodes. Note that most, if not all, of these nodes must have been already indexed within the indexing process of the previously visited nodes or different/new base suffix under node $a$, hence, what is actually needed to be indexed is the label between the leaf node of base suffix $x$ and its parent (tail of base suffix). The parent node of the leaf node of base suffix $x$ or other nodes may need to be indexed (caused mainly by different/new base suffix under node $a$). These nodes that need to be indexed can be detected with minor costs using the same rules used to find base paths (Section 5). The following code snapshot shows the full needed code to $OT$ index base suffixes non-trivially by indexing the tails of base suffixes. 

\begin{lstlisting}[language=Python, caption= OT indexing function using base suffixes (non-trivial algorithm by indexing tails of base suffixes)]
def phase_2_for_OT_indexing_for_base_paths(tree, k):
		# find base paths, record base and bottom nodes, and create OT index
		stack.append(tree.root)
		children_stack.append((list(tree.root.nodes_link_to_me)))
		key_stack.append(0)


		while stack:
			current_node = stack[-1]	
			# check if OSHR[current_node.key] is empty, then remove it from stack
			if len(children_stack[-1]) > 0:
				last_node_under_top_node_in_stack = children_stack[-1][-1]
				stack.append(last_node_under_top_node_in_stack)		
				children_stack[-1].pop()
				if hasattr(last_node_under_top_node_in_stack, "nodes_link_to_me"):
					children_stack.append((list(last_node_under_top_node_in_stack.nodes_link_to_me)))
				else:
					children_stack.append([])

				key_stack.append(tree.keys_counter + 1)	 
				   
			else:
								
				for base_suffix in current_node.base_suffixes:
					suffix_idx = base_suffix - current_node.depth
					leaf_node = tree.leaf_suffix_index_to_leaf_memory_list[suffix_idx]
					node = tree.leaf_suffix_index_to_leaf_memory_list[base_suffix]
					req_depth = leaf_node.parent.depth - current_node.depth 
					transition_letter =  text[node.idx + current_node.depth:node.idx + current_node.depth + k]
					
					if  hasattr(current_node, "nodes_link_to_me"):
						while node.depth >= req_depth and tree.root != node:
							tree.keys_counter += 1
							mapping_guided_suffix = base_suffix
							OT_indx = tree.keys_counter
							
							node.OT_indexes.append((OT_indx, transition_letter))
							node = node.parent	
					else:
						while node.depth > 0:
							tree.keys_counter += 1
							mapping_guided_suffix = base_suffix
							OT_indx = tree.keys_counter
							
							node.OT_indexes.append((OT_indx, transition_letter))
							node = node.parent	
						
				if hasattr(current_node, "left_OT_index"):
					current_node.left_OT_index = key_stack[-1]
					current_node.right_OT_index = tree.keys_counter
				else:
					setattr(current_node, "left_OT_index", defaultdict())
					setattr(current_node, "right_OT_index", defaultdict())
					
					current_node.left_OT_index = key_stack[-1]
					current_node.right_OT_index = tree.keys_counter	
				key_stack.pop()
				stack.pop()
				children_stack.pop()
				
\end{lstlisting}

\subsection{Searching process}

The following algorithm shows how to search for patterns with up to $k$ Hamming distance (for edit distance, it can be applied with minor tweaks).

In this section we will describe searching for $k = 1$. Once the paths in the tree that are the results of $k = 1$ matching are found, we will use these paths and repeat the process (for $k= 1$) to find the results for $k = 2$, and so forth. 

Firstly, find the last node that was reached by walking each suffix in the pattern in $ST$. If a mismatch occurred in the middle of an edge, skip it, record this mismatch, and keep walking. If walking ends in the middle of an edge, then return the sink node of that edge. This process will cost linear time and space using the suffix links in $ST$. Moreover, if the end node of walking a suffix is node $a$ and the label of suffix is $S$, then the end nodes of suffix $\sigma S$ for each possible $\sigma$ in $\Sigma$ must be the nodes that have suffix links to node $a$. 

Now, walk with the pattern in $ST$. If the walk is on an edge and a mismatch occurred at position $x$, then proceed the walking as exact matching until the end of the pattern is reached. If exact-matching walking ends at a node $a$, report suffix indexes that are under node $a$ as the approximate matching results for position $x$ which. If not, report no approximate matching with $k=1$ value.

If the walk encountered no mismatches on an edge and reached an internal node, let's say node $a$, then find (in $O(log_2n)$ time) the positions of $left\_OT\_index$ and $right\_OT\_index$ in $OT\_indexes$ list of node $t_\sigma$ where $t_\sigma$ is the end node of walking string $\sigma p_d p_{d+1} p_{d+2}...p_m$ in $ST$, $p_i$ is the letter at position $i$ in the pattern, $d$ is $depth$ of node $a$, and $\sigma$ is a letter in $\Sigma$ (not equal to $p_d$). Now, if there are \textit{OT} index values inbetween the \textit{found} positions in the $OT\_indexes$ list, this mean there must be a path under node $a$ for string $\sigma p_d p_{d+1} p_{d+2}...p_m$ and the occurrences will be the suffix indexes associated with each of these values (with extra constant computation for each suffix index, the association is extracted from $OT\_index$ list/dictionary). If there is not, then there is no path under node $a$ with a label equal to string $\sigma p_d p_{d+1} p_{d+2}...p_m$. 

Repeat the same process for each $\sigma$ in $\Sigma$ and each internal node encountered in the path of the pattern in $ST$.

\subsection{\textit{OT} indexing using base path}
The algorithm starts by traversing the $OSHR$ tree in postorder then at each visited node find the bottom nodes, according to the algorithm in Section 5 for finding base paths. The visited node (considered here as the top node for base paths) with each found bottom node forms the set of the base paths under the visited node. 

$OT$ indexing of base suffixes involved walking in $ST$ explicitly in order to map base suffixes with their last extent suffix. This way, $OT$ indexes were directly added to the node memory. For base paths indexing, mapping each base path to its last extent path (the one where the top node is the tree root) by explicit walking in $ST$ would cost more than constant time especially if there are several nodes (up to $h$ nodes) between the top node and the bottom node of the base path. Therefore, an indirect/tricky mapping procedures were performed, shown in the code snapshot in both listing 3 and 7, in order to achieve linear time mapping costs.

\begin{lstlisting}[language=Python, caption= Mapping procedure for base paths]
def phase_3_for_OT_indexing_for_base_paths(tree):
		#map OT indexes of base paths to same path starting from root of ST, then sort OT indexes at node
		
		#iterative processings
		nodes_stack.append(tree.root)
		children_stack.append((list(tree.root.transition_links[x] for x in sorted(tree.root.transition_links.keys(), reverse=True))))
		
		temp = defaultdict(int)
		sum_of_all_OT_indexes = 0
		
		while nodes_stack:
			current_node = nodes_stack[-1]				  
			if len(children_stack[-1]) > 0:
				last_node_under_top_node_in_stack = children_stack[-1][-1]
				#iterative processings
				nodes_stack.append(last_node_under_top_node_in_stack)	  # append it to process later with the required order (postorder) and remove it from OSHR[current_node.key]
				children_stack[-1].pop()					
				children_stack.append((list(last_node_under_top_node_in_stack.transition_links[x] for x in sorted(last_node_under_top_node_in_stack.transition_links.keys(), reverse=True))))
			else:
				# alongside processings
				if not current_node.is_leaf():
					setattr(current_node, "OT_indexes", [])
					if current_node.depth in tree.temp_dict:
						#print (tree.temp_dict[current_node.depth])
						for i in range(len(tree.temp_dict[current_node.depth])-1, -1, -1):
							suffix_idx = tree.temp_dict[current_node.depth][i][0]
							OT_indx = tree.temp_dict[current_node.depth][i][1]
							transition_letter = tree.temp_dict[current_node.depth][i][2]
						
							if current_node.index_of_leftmost_leaf <= suffix_idx <= current_node.index_of_rightmost_leaf:
								current_node.OT_indexes.append((OT_indx, transition_letter))
								tree.temp_dict[current_node.depth].pop()
							else:
								break
								
					# now sort OT indexes for each transition letters
					current_node.OT_indexes.sort()
					sum_of_all_OT_indexes += len(current_node.OT_indexes)
					#print (current_node.OT_indexes)
					
					# index transition_letters based on their positions in OT_indexes. No need to sort the list as it's already sorted
					setattr(current_node, "transition_letters_position_in_OT_indexes", defaultdict(list))
					for i in range(len(current_node.OT_indexes)):
						OT_index = current_node.OT_indexes[i][0]
						transition_letters = current_node.OT_indexes[i][1]
						
						current_node.transition_letters_position_in_OT_indexes[transition_letters].append(i)
						current_node.OT_indexes[i] = OT_index
					
				#iterative processing 
				nodes_stack.pop()
				children_stack.pop()
	
			
	nodes_stack = []
	children_stack = []
	start = time.time()
	phase_3_for_OT_indexing_for_base_paths(tree)

\end{lstlisting}

As the \textit{OT} indexing process of base paths involves indexing only internal nodes and involving the indexing of tails label (label/string between leaf nodes and their parents), the search process is as follows. If the search process encounters an internal node, let's say node $a$, then using the code described below, find the approximate matching between node $a$ and each of its child leaf node where the cost will be constant time for each leaf node (as shown in the code below). As there are $O(\Sigma)$ leaf nodes under any internal node, then the cost will be $O(\Sigma)$. 

\begin{lstlisting}[language=Python, caption=Matching label of leaf nodes]
for node in reached_node.transition_links.values():
	if node.is_leaf():
		suffix_number_under_node = tree.leaf_suffix_index_to_leaf_memory_list[node.idx + reached_node.depth]
		end_node_of_suffix_starting_from_root = suffixes_traversals[reached_node.depth][-1][0]
		if end_node_of_suffix_starting_from_root.is_leaf():
			if suffix_number_under_node.idx == end_node_of_suffix_starting_from_root.idx:
				tree.matching_nodes.append(node)
				print ("Found second order match for", transition_letters)
		else:
			if suffix_number_under_node.key  >= end_node_of_suffix_starting_from_root.index_of_leftmost_leaf and suffix_number_under_node.key <= end_node_of_suffix_starting_from_root.index_of_rightmost_leaf:
					tree.matching_nodes.append(node)
\end{lstlisting}

Before describing how to find approximate matching between node $a$ and any of its descendant internal nodes, let's define a leaf node that has at least one sibling internal node to be referred as uncle leaf node and the one that all of its siblings are leaf nodes to be referred as non-uncle leaf node. Clearly, uncle leaf node must be child node of grand internal node whereas non-uncle one must be a child node of non-grand node. 

Now, if the walk encountered no mismatches on an edge and reached an internal node, let's say node $a$, then find (in $O(log_2n)$ time) the positions of $left\_OT\_index$ and $right\_OT\_index$ in $OT\_indexes$ list of node $t_\sigma$ where $t_\sigma$ is the end node of walking string $\sigma p_d p_{d+1} p_{d+2}...p_m$ in $ST$. If there are \textit{OT} index values inbetween the \textit{found} positions in $OT\_indexes$ list, this means there must be a path ends at an internal node, let's say node $x$, under node $a$ where the label of the path match string $\sigma p_d p_{d+1} p_{d+2}...p_m$. As a result, walk to node $x$ (with $O(log_\Sigma)$ cost not $O(m)$ using a guided leaf node as shown in the code), then once reached node $x$ perform the matching between string $p_j p_{j+1} p_{j+2}...p_m$ and the edges between node $x$ and the each of its child nodes (costing $O(\Sigma)$, where $j$ is the depth of node $x$). If there is no \textit{OT} index values, this means there is no path ends at an \textit{internal node} under node $x$ with depth equal to the depth of node $t_\sigma$. Due to this a backtracking process must be performed in order to search for matching between the parent nodes of node $t_\sigma$ and their leaf nodes (that must be uncle leaf nodes). So, repeat the same process with the parent node of $t_\sigma$. Keep repeating the same process until an occurrence is found or the root node is reached. This backtracking process will add an additional cost of $O(log_\Sigma n)$ (which is the expected number of backtracking times).  

Repeat the same process for each $\sigma$ in $\Sigma$ and each internal node encountered in the path of pattern in $ST$.

\subsection{\textit{OT} indexing using base uncle suffixes}
In order to avoid the cost caused by $O(log_\Sigma n)$ factor when \textit{OT} indexing base paths, indexing of base paths and \textit{base uncle suffixes} is sought. Base uncle suffixes are the first appearance, throughout postorder traversal of $OSHR$ tree, of a suffix where the leaf node of this suffix is an uncle leaf node. Theoretically, the cost of \textit{OT} indexing base paths and base uncle suffixes will be $O(nlog_\Sigma n)$. However, note that indexing base uncle suffixes will implicitly cover most of base paths so there is no need to index base paths when indexing base uncle suffixes. Moreover, the practical cost of $OT$ indexing base uncle suffixes is near-linear. Hence, $OT$ indexing of base uncle suffixes can provide almost linear solution similar to $OT$ indexing of base suffixes and search costs without the $O(log_\Sigma n)$ ($O(m + \frac {log_2n {(log_\Sigma n)} ^{k}}{k!} + occ)$  which is the same as $OT$ indexing of base suffixes). 

In order to find base uncle suffixes and achieve this with linear costs, base suffixes can be used. The following code snapshot provide a linear algorithm to perform so. 

\begin{lstlisting}[language=Python, caption=Finding Uncle base suffixes]
def find_base_uncle_suffixes(tree):	
		# find base suffixes
		stack.append(tree.root)
		children_stack.append((list(tree.root.transition_links[x] for x in sorted(tree.root.transition_links.keys(), reverse=True))))		
		cost = 0
		setattr(tree, "singleton_suffixes", defaultdict(int))
		
		while stack:
			current_node = stack[-1]				  
			if len(children_stack[-1]) > 0:
				last_node_under_top_node_in_stack = children_stack[-1][-1]
				stack.append(last_node_under_top_node_in_stack)	 
				
				children_stack[-1].pop()					
				children_stack.append((list(last_node_under_top_node_in_stack.transition_links[x] for x in sorted(last_node_under_top_node_in_stack.transition_links.keys(), reverse=True))))										   
			else:
				stack.pop()
				children_stack.pop()
				
				# alongside processings
				if not current_node.is_leaf() :
					for base_suffix in current_node.base_suffixes:
						suffix_idx = base_suffix - current_node.depth
						leaf_node = tree.leaf_suffix_index_to_leaf_memory_list[suffix_idx]
						if leaf_node.parent.list_of_deepest_inetrnal_node_for_all_paths[-1][0] != -1 and leaf_node.parent != current_node:
							current_node.base_uncle_suffixes.append(base_suffix)
							tree.singleton_suffixes[base_suffix] = 0
							cost += 1
						else :
							leaf_with_next_suffix = tree.leaf_suffix_index_to_leaf_memory_list[suffix_idx + 1]
							next_top_base = current_node._suffix_link
							next_suffix_index = leaf_with_next_suffix.idx
							
							while True:
								cost += 1
								if leaf_with_next_suffix.parent.list_of_deepest_inetrnal_node_for_all_paths[-1][0] != -1 and leaf_with_next_suffix.parent != next_top_base:
									next_top_base.base_uncle_suffixes.append(next_suffix_index + next_top_base.depth)
									tree.singleton_suffixes[next_suffix_index + next_top_base.depth] = 0
									break
				
								elif next_top_base == tree.root:
									break
									
								else:
									leaf_with_next_suffix = tree.leaf_suffix_index_to_leaf_memory_list[next_suffix_index + 1]
									next_top_base = next_top_base._suffix_link
									next_suffix_index = leaf_with_next_suffix.idx
								
		print ("Finding base uncle suffixes took", cost)	

	stack = []
	children_stack = []
	start = time.time()
	find_base_uncle_suffixes(tree)
\end{lstlisting}

The expected number of base uncle suffixes can be $n$ or less as there are suffixes which never become uncle suffixes. Once all base uncle suffixes determined, find (if any) suffix indexes that were not as such (let this set denoted as $A$). Now, $OT$ indexing of set $A$ (fully not only tails) along with $OT$ indexing of only the tails of base uncle suffixes, as shown in the following code snapshot, will provide a near-linear index of $ST$ with searching cost of $O(m + \frac {log_2n {(log_\Sigma n)} ^{k}}{k!} + occ)$. 

\begin{lstlisting}[language=Python, caption=OT indexing of uncle base suffixes]
def phase_3_for_OT_indexing_for_base_paths(tree, k):
	stack.append(tree.root)
	children_stack.append((list(tree.root.nodes_link_to_me)))
	key_stack.append(0)

	
	
	temp_leaf = defaultdict(int)
	temp_internal = defaultdict(int)
	temp_keys = defaultdict(int)
	
	setattr(tree, "temp_dict", defaultdict(list))
	already_indexed_suffixes = defaultdict(int)
		
	while stack:
		current_node = stack[-1]	
		# check if OSHR[current_node.key] is empty, then remove it from stack
		if len(children_stack[-1]) > 0:
			last_node_under_top_node_in_stack = children_stack[-1][-1]
			stack.append(last_node_under_top_node_in_stack)		
			children_stack[-1].pop()
			if hasattr(last_node_under_top_node_in_stack, "nodes_link_to_me"):
				children_stack.append((list(last_node_under_top_node_in_stack.nodes_link_to_me)))
			else:
				children_stack.append([])

			key_stack.append(tree.keys_counter + 1)	 
			   
		else:
			for base_suffix in current_node.base_uncle_suffixes:
				suffix_idx = base_suffix - current_node.depth
				leaf_node = tree.leaf_suffix_index_to_leaf_memory_list[suffix_idx]
			
				node = tree.leaf_suffix_index_to_leaf_memory_list[base_suffix]
				req_depth = leaf_node.parent.depth - current_node.depth 
				transition_letter =  text[node.idx + current_node.depth:node.idx + current_node.depth + k]

				while node.depth > req_depth:
					tree.keys_counter += 1
					mapping_guided_suffix = base_suffix
					OT_indx = tree.keys_counter
					
					node.OT_indexes.append((OT_indx, transition_letter))
					if node.is_leaf():
						temp_leaf[tree._edgeLabel(node, tree.root)] += 1
					else:
						temp_internal[tree._edgeLabel(node, tree.root)] += 1
					temp_keys[str(base_suffix) + "-" + str(node.key) + "-" + str(current_node.key)] = 1
					node = node.parent	
			
			
			for base_suffix in current_node.base_suffixes:
				if base_suffix not in tree.singleton_suffixes:
					suffix_idx = base_suffix - current_node.depth
					node = tree.leaf_suffix_index_to_leaf_memory_list[base_suffix]
					transition_letter =  text[node.idx + current_node.depth:node.idx + current_node.depth + k]

					while node.depth > 0:
						tree.keys_counter += 1
						mapping_guided_suffix = base_suffix
						OT_indx = tree.keys_counter
						
						node.OT_indexes.append((OT_indx, transition_letter))
						if node.is_leaf():
							temp_leaf[tree._edgeLabel(node, tree.root)] += 1
						else:
							temp_internal[tree._edgeLabel(node, tree.root)] += 1
						temp_keys[str(base_suffix) + "-" + str(node.key) + "-" + str(current_node.key)] = 1
						node = node.parent	
							
			
			if hasattr(current_node, "left_OT_index"):
				current_node.left_OT_index = key_stack[-1]
				current_node.right_OT_index = tree.keys_counter
			else:
				setattr(current_node, "left_OT_index", defaultdict())
				setattr(current_node, "right_OT_index", defaultdict())
				
				current_node.left_OT_index = key_stack[-1]
				current_node.right_OT_index = tree.keys_counter	
			key_stack.pop()
			stack.pop()
			children_stack.pop()

stack = []
children_stack = []
key_stack = []
tree.keys_counter = -1

setattr(tree, "OT_index", defaultdict())

phase_3_for_OT_indexing_for_base_paths(tree, K)
\end{lstlisting}

\end{document}